\documentstyle[aps]{revtex}

\begin{document}

\draft

\title{
No-cloning and superluminal signaling
}

\author{S.J. van Enk}
\address{Norman Bridge Laboratory of Physics\\
California Institute of Technology 12-33\\
Pasadena CA 91125, USA}
\date{March 12, 1998}
\maketitle

\begin{abstract}
The argument given by M.D. Westmoreland and B. Schumacher in quant-ph/9801014 leading to the purported conclusion that superluminal signaling violates the no-cloning theorem is shown to be incorrect. 
\end{abstract}


There seems to be no reason why standard nonrelativistic quantum mechanics should be inconsistent with the superluminal transmission of classical information. Nevertheless, a recent paper \cite{west} claims that the no-cloning theorem \cite{noclon}---a simple consequence of the linearity of quantum mechanics---is violated
if superluminal signals are allowed in the teleportation protocol of \cite{bennett}, a protocol which uses only standard nonrelativistic quantum mechanics.
Although the authors of \cite{west} invoke special relativity, they emphasize that 
the contradiction they arrive at is not the usual causal paradox that is found by combining relativity with superluminal signaling.
On the other hand, given that also the combination of special relativity and standard quantum mechanics leads to well-known problems \cite{ahar}, inconsistencies seem unavoidable. 

The teleportation protocol \cite{bennett} utilizes an EPR pair and classical communication to teleport an arbitrary quantum state  $|\psi\rangle$ from one party (Alice) to another (Bob), where the state $|\psi\rangle$ is unknown to both parties.
Alice performs a joint measurement on the two particles on her side, one containing the state $|\psi\rangle$, the other constituting half of an EPR pair shared by her and Bob. She communicates the measurement outcome to Bob, after which he can use this information to apply the correct unitary operation to his part of the EPR pair. The resultant is that his particle ends up in the state $|\psi\rangle$. 

The argument of \cite{west} is now as follows:
If the communication between Alice and Bob is allowed to be superluminal, then Alice having the state $|\psi\rangle$ on her side and Bob obtaining the state $|\psi\rangle$ on his are separated by a spacelike interval. Thus there is a Lorentz frame ${\cal L}$ in which the order of events is such that Bob obtains the state $|\psi\rangle$ {\em before} Alice performs her joint measurement, and hence {\em before} Alice loses her copy of $|\psi\rangle$.  In that frame ${\cal L}$, until the moment Alice performs her measurement, both Alice and Bob have a copy of the unknown state $|\psi\rangle$, in contradiction with the no-cloning theorem \cite{noclon}.

However, upon examining the argument in more detail it soon becomes clear that there are in fact never two copies of the same state in any Lorentz frame. To be sure, there will still be inconsistencies, but those are due to the attempt to combine relativity with superluminal signaling. We describe the observations as done by two observers from two different Lorentz frames.  

One could in principle attempt to treat the whole experiment, including Alice and Bob, quantum mechanically.  However, it is hard to see how to describe a superluminal signal consistently within a relativistic quantum-mechanical framework. Therefore,
as in \cite{west}, we treat Alice, Bob and their measurement devices as classical and describe measurements as standard Von Neumann measurements of the first kind.

First consider the description of the teleportation experiment by observer Carol, who is at rest relative to Alice and Bob, and who actually provided Alice with the state $|\psi\rangle$.
According to Carol, the joint initial state of Alice's particles $A_1$ and $A$, and Bob's particle $B$ can be written as
\begin{equation}
\label{one}
|\Psi\rangle_{A_1,A,B}=|\psi\rangle_{A_1}|{\rm EPR}\rangle_{A,B},
\end{equation}
where $|{\rm EPR}\rangle$ stands for a maximally entangled state.
If for simplicity we assume the particles to be two-state quantum systems, this state can be {\em rewritten} as
\begin{equation}
\label{onep}
|\Psi\rangle_{A_1,A,B}=\frac{1}{2}\sum_{i=1}^4 |{\rm EPR}_i\rangle_{A_1,A} U_i|\psi\rangle_B,
\end{equation}
in terms of the standard four Bell states and four corresponding unitary operators $U_i$ \cite{bennett}.
Alice then performs a measurement that distinguishes between the four Bell states and loudly declares the outcome so that even Carol can hear it. If the outcome of that measurement is $i=i_0$, the state, according to Carol, is collapsed into
\begin{equation}\label{two}
|\Psi\rangle_{A_1,A,B}\mapsto|{\rm EPR}_{i_0}\rangle_{A_1,A} U_{i_0}|\psi\rangle_B.
\end{equation}
Subsequently, Alice tells Bob her result (using superluminal signaling), after which he  applies the unitary operation $U^{\dagger}_{i_0}$ to obtain
\begin{equation}
\label{three}
|\Psi\rangle_{A_1,A,B}\mapsto|{\rm EPR}_{i_0}\rangle_{A_1,A}|\psi\rangle_B,
\end{equation}
which completes the teleportation process: Bob now possesses the state $|\psi\rangle$.

Now we introduce Carol's colleague, Dennis, who is moving with a velocity close to $c$ such that the time order of Bob applying his unitary operation and Alice performing her measurement is reversed as compared to Carol's view. Suppose moreover that Carol has revealed to him the state $|\psi\rangle$.
Ignoring for simplicity the particular transformation of the quantum states and of unitary operators \cite{sakurai} due to the Lorentz transformation, Dennis's description of the initial state is the same as for Carol, namely as given by (\ref{one}) or (\ref{onep}).

Dennis then observes that Bob applies the unitary operation $U^{\dagger}_{i_0}$. Dennis now assigns the state
\begin{equation}
\label{twop}
\frac{1}{2}\sum_{i=1}^4|{\rm EPR}_{i}\rangle_{A_1,A} U^{\dagger}_{i_0}U_{i}|\psi\rangle_B=|\psi\rangle_{A_1}U^{\dagger}_{i_0}|{\rm EPR}\rangle_{A,B}
\end{equation}
to Alice's and Bob's particles. This leaves Alice, but not Bob, with a particle in state $|\psi\rangle$.
Subsequently Alice performs her measurement, and manages to get the result $i=i_0$, which collapses the state to (\ref{three}), completing teleportation.  
The main point to note, of course, is that at all stages there is only a single copy of $|\psi\rangle$.

There is here the usual causal paradox that Bob applies the correct unitary operation before Alice gets the corresponding measurement outcome. Or, put in a different way, although Alice has only 25\% probability of finding the result $i=i_0$ according to Dennis's description (\ref{twop}), in reality she will {\em always} find that result. 
Similarly, if Bob were to do a measurement to confirm whether he has the state $|\psi\rangle$ or not, he will {\em always} find the answer "yes" (the only answer consistent with the future measurement outcome $i=i_0$), although Dennis would assign only a probability of 50\% to that result.  

Also note that Bob cannot design this measurement by himself, since he has no knowledge about the state $|\psi\rangle$. Thus, either Carol or Dennis would have to prepare a measuring device for him. This means that, although after the confirmation measurement both Alice and Bob would possess a copy of $|\psi\rangle$ (according to Dennis), this would {\em not} be a violation of the no-cloning theorem, since only cloning of {\em unknown} quantum states is forbidden.

The conclusion that superluminal signaling violates the no-cloning theorem is not warranted. The naive application of Lorentz transformations to (imprecise) statements of the form ``Alice has the state $|\psi\rangle$'' is incorrect.
Although the combination of special relativity, standard quantum mechanics and superluminal communication leads to inconsistencies, the conclusion that, therefore, the latter two are inconsistent is hard to justify, given the problems of combining special relativity both with standard quantum mechanics and with superluminal signaling in the first place.

It is a pleasure to thank Chris Fuchs for many useful comments and entertaining discussions.

\end{document}